\documentclass[
aps,%
12pt,%
final,%
notitlepage,%
oneside,%
onecolumn,%
nobibnotes,%
nofootinbib,% В текущей версии REVTeX c этой опцией
superscriptaddress,%
noshowpacs,%
centertags]%
{revtex4-1}

\usepackage{appendix}
\usepackage{dcolumn}
\usepackage{amsmath,amssymb}
\usepackage{graphicx}
\usepackage{float}
\usepackage{bigints}
\usepackage{natbib}
\usepackage{xcolor}
\usepackage{subcaption}

\begin{document}

\title{Proposition of FSR Photon Suppression Employing a Two-Positron Decay Dark Matter Model to Explain Positron Anomaly in Cosmic Rays}

	\author{\firstname{Ramin}~\surname{Barak}}
	
        \email{Correspondence: ramin.k.barak@gmail.com}

        \affiliation{%
		National Research Nuclear University MEPhI \\
		Kashirskoe highway 31, Moscow, 115409
	}

        \affiliation{%
        Joint Institute for Nuclear Research (JINR)\\
        Dubna, Russia, 141980
        }
        
        \author{\firstname{Konstantin}~\surname{Belotsky}}

        \affiliation{%
		National Research Nuclear University MEPhI \\
		Kashirskoe highway 31, Moscow, 115409
	}

        \affiliation{%
		Novosibirsk State University \\
		Pirogova str. 1, Novosibirsk, Russia, 630090
	}

        \author{\firstname{Ekaterina}~\surname{Shlepkina}}
	
        %\affiliation{
	%	National Research Nuclear University MEPhI \\
	%	Kashirskoe shosse 31, Moscow, 115409
	%}
	
	%\author{\firstname{Konstantin}~\surname{Belotsky}}
	%\email{E-mail: koroaa@jinr.ru}
	\affiliation{%
		National Research Nuclear University MEPhI \\
		Kashirskoe highway 31, Moscow, 115409
	}

	%\email{E-mail: koroaa@jinr.ru}
	%\affiliation{%
	%	National Research Nuclear University MEPhI \\
	%	Kashirskoe highway 31, Moscow, 115409
	%}
	
\begin{abstract}
The origin of an anomalous excess of high-energy (about 100 GeV and higher) positrons in cosmic rays is one of the rare problems in this field, which is proposed to be solved with dark matter (DM). Attempts to solve this problem are faced with the issue of having to satisfy the data on cosmic positrons and cosmic gamma radiation, which inevitably accompanies positron production, such as FSR (final state radiation), simultaneously. We have been trying to come up with a solution by means of two approaches: making assumptions (*) about the spatial distribution of the dark matter and (**) about the physics of its interactions. This work is some small final step of a big investigation regarding the search for gamma suppression by employing the second approach, and a model with a doubly charged particle decaying into two positrons ($X^{++}\rightarrow e^+e^+$) is suggested as the most prospective one from those considered before. 
\end{abstract}


	\maketitle
 \newpage

\section{Introduction}
The physical nature of dark matter (DM) is the subject of long-term investigations. Different sophisticated research methods have been elaborated. Among them, there are indirect ones concerning a possible explanation of cosmic ray (CR) anomalies. 
%Different possibilities can be suggested to explanation of CR positron anomaly \cite{[]} is considered. 
Cosmic positrons manifest anomalous growth in the energy spectrum in the range of 10--500 GeV, as observed by PAMELA \cite{Ref1}, AMS-2 \cite{Ref2}, and Fermi \cite{Ref3}, and possibly at higher energies, as pointed out by, e.g., DAMPE \cite{Ref4} (positron anomaly (PA)). Basically, two following explanations are suggested: the ones related to pulsars \cite{Ref5, Ref6, Ref7, Ref8} and the ones related to the annihilation or decay of DM particles (see, e.g., \cite{ Ref9, Ref10, Ref11, Ref12, Ref13, Ref14, Ref15, Ref16, Ref17, Ref18, Ref19}). There have also been attempts, based on supernova explosions \cite{Ref20, Ref21}, changes of a CR propagation model \cite{Ref22, Ref23, Ref24}, and some others. However, all of these at least suffer from the problem of fine-tuning of model parameter magnitudes. 

\begin{figure}[H]
\begin{center}
\begin{minipage}[h]{0.496\linewidth}
\includegraphics[width=1\linewidth]{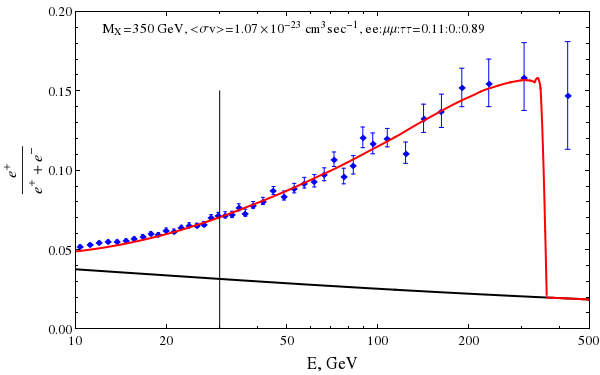}
\label{constr_10} %% метка рисунка для ссылки на него
\end{minipage}
\hfill
\begin{minipage}[h]{0.496\linewidth}
\includegraphics[width=1\linewidth]{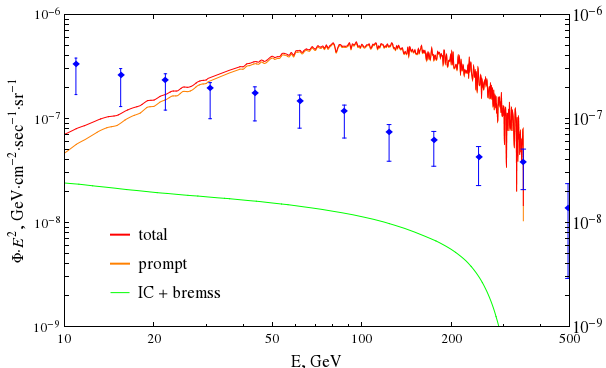}
\label{constr_01}
\end{minipage}
\end{center}
\caption{A comparison of the data of CR experiments and the predicted results of the model for decaying DM: (\textbf{left}) cosmic positron fraction, red line indicates the theoretical prediction, black line is the expected background, and data points are from AMS-02; (\textbf{right}) IGRB, red line corresponds to the total expected contribution of photons of the same DM model as in the left plot, and the data points are of the Fermi-LAT. Figures were taken from \cite{Ref11}}
\label{fig1}
\end{figure}

Here, we are trying to reduce the fine-tuning problem in the framework of a DM explanation of PA. This explanation faces the issue of disagreement with data on cosmic gamma radiation, first of all, the so-called isotropic gamma-ray background (IGRB) obtained by Fermi-LAT \cite{Ref11}; an illustration is provided in Figure \ref{fig1}. The authors are aware that Figure \ref{fig1} does not contain the latest data from the AMS experiment \cite{Ref25}. However, the choice for this particular figure was made nonetheless in order to portray the issue, which only intensifies in case of an increase in energy range as in new AMS data. Any positrons (electrons) $e^+e^-$ produced by annihilation or decay of DM particles induce prompt photons (mainly final state radiation (FSR)) and photons due to the interaction of $e^+e^-$ with medium photons (mainly due to inverse Compton (IC) scattering on starlight). As one can see from Figure \ref{fig1}, the main problem arises due to, basically, FSR photons and occurs at high energies.

{Generally, in spite of this work initially being connected to the problems of dark matter and positron anomaly, it does not exclude its independent importance concerning the issue of FSR suppression.}

\section{Approaches to the Positron Anomaly Solution with Dark Matter}

It is possible to propose two approaches for solving the problem of disagreement with gamma-ray data in a DM explanation of PA. First, one is due to a {spatial} distribution of DM components, including DM clumps and other structures, such as dark disk. The second approach is related to the physical properties of DM particles that govern the \mbox{decay/annihilation process.}

Our group proposed the so-called ``dark disk model'' \cite{Ref11, Ref26, Ref27, Ref28, Ref29, Ref30, Ref31} in the framework of the first approach in order to explain positron anomaly in AMS-02 data. The idea is the following. {The contradiction is caused by a finite traveling length of high-energy positrons because of the energy losses they suffer and the existence of a magnetic field around the galactic disk, which makes their trajectory tangled and does not allow positrons born outside of it to reach the Earth. However, gammas are unaffected by these and, therefore, arrive from virtually all the distances.} %contribute to the total gamma-ray flux. 
This enables one to artificially decrease the amount of gamma flux while keeping the amount of positrons unchanged by ``cutting off'' an area of space outside the magnetic disk. In fact, there can be one minor ``active'' component of DM that gives a positron signal and a major passive one, which forms a halo of the galaxy. It was shown in \cite{Ref10} that the implementation of this particular model greatly reduces the contradiction with IGRB data. 

In the framework of the second approach, different attempts were undertaken to find a physical model of DM (Lagrangian) to provide the suppression of gamma-ray output. However, the focus here lies on doubly charged DM particles.

Earlier, DM models based on technicolor \cite{Ref32, Ref33}, where doubly charged techniparticles, in the composition of {electrically neutral} dark atoms, decay into two positrons ($X^{++}\rightarrow e^+e^+$), were considered. Details of a technicolor DM model can be found in {the Appendix} \ref{Appendix: A}. 

Additionally, different DM models with doubly charged particles, based on various standard model extensions \cite{Ref34, Ref35, Ref36, Ref37, Ref38}, were discussed and elaborated to solve the contradiction of the results of an underground experiment DAMA with the results of other similar experiments. {We do not fix any specific physical model here, but it is believed that such DM candidate containing a doubly charged particle, hidden inside compact neutral dark atoms, \footnote{\footnotesize The size and binding energy of such an atom $\{X^{++}Y^{--}\}$ is defined by the mass value of each component, which is, as a rule, required to be $\gtrsim 1$ TeV for PA explanation.} may avoid different constraints of underground, (above)ground, and space experiments (see, e.g. \cite{Ref39}), including those based on the observation of white dwarfs \cite{Ref40}.} As to the positron anomaly, a model with the decay $X^{++}\rightarrow e^+e^+$ has a simple advantage as compared with the more traditional one $X^0\rightarrow e^+e^-$, since there are twice as many positrons per one FSR photon. 

In this short letter, we follow the second approach related to the physical properties of DM, which account for decay with positron production. More specifically, our aim was to point out that the DM model with a doubly charged unstable particle has one more advantage in the context of positron anomaly solution. This additional advantage is associated with two identical particles in the final state \cite{Ref10}. Such a system ($e^+e^+$) does not have classical dipole radiation since it has a zero electric dipole moment. The so-called ''single photon theorem'' (or ''radiation zeros'') \cite{Ref41} claims partial suppression of identically charged particle radiation, thus restoring a correspondence between classical and quantum descriptions to some extent. Here, we demonstrate a possible role of this suppression in relation to the physics of dark matter in explaining the cosmic positron anomaly.

\section{Models Used}

Following the %theoretical considerations in \cite{ref-journal11} and 
theoretical simplification of the model in \cite{Ref10}, two models of a decaying dark matter particle were considered.

\begin{itemize}
\item A model with a decay of a scalar DM particle into two positrons
\begin{equation}
X^{++} \rightarrow e^+ + e^+ 
\end{equation}
according to Lagrangian
\begin{equation}
    L_{int}=X\bar\Psi^C(a+b\gamma_5)\Psi+h.c.
\end{equation}
with an accompanying decay of a DM particle into two positrons and an FSR photon
\begin{equation}
X^{++} \rightarrow e^+ + e^+ + \gamma ;
\end{equation}
\item and a more conventional model, to be compared with, with a decay of a scalar DM particle into an electron and a positron
\begin{equation}
X^0 \rightarrow e^+ + e^- 
\end{equation}
according to Lagrangian
\begin{equation}
    L_{int}=X\bar\Psi(a+b\gamma_5)\Psi+h.c.
\end{equation}
respectively accompanied by a decay
\begin{equation}
X^0 \rightarrow e^+ + e^- + \gamma .
\end{equation}
\end{itemize}
$\Psi$ represents the positron/electron wave function, index $C$ stands for charge conjugation, $a=b=1$ was used in this work during calculations, and $\gamma_5$ is the Dirac matrix. 

%\textbf{Lagrangians ...(it's better to place it right after the corresponding process (?) (ES)).}

%textbf {This one is for $X \rightarrow e^+ + e^- + \gamma$ (and $X \rightarrow e^+ + e^-$). Should we delete parameterization from Lagrangian formulas?}

%\begin{equation}
%\mathscr{L} = X\Bar{\psi}^C(a+b\gamma^5)\psi + X^*\Bar{\psi}(a-b\gamma^5)\psi^C + %\Bar{\psi}\gamma^{\mu}A_{\mu}\psi.
%\end{equation}

%\textbf {This one is for $X \rightarrow e^+ + e^+ + \gamma$ (and $X \rightarrow e^+ + e^+$ of course)}

%\begin{equation}
%\mathscr{L} = X\Bar{\psi}(a+b\gamma^5)\psi + X\Bar{\psi}(a-b\gamma^5)\psi + %\Bar{\psi}\gamma^{\mu}A_{\mu}\psi .
%\end{equation}

Photon (FSR) suppression is of interest to us, since it is necessary to eliminate the contradiction with the excess of IGRB during the decay of DM particles. This implies that the ratio of the width of the three-body to two-body decay should be minimal: 

\newpage
\begin{equation}
\frac{\Gamma(X \rightarrow e^+e^{\pm}\gamma)}{\Gamma(X \rightarrow e^+e^{\pm})}\equiv Br(X \rightarrow e^+e^{\pm}\gamma) =min.
\end{equation}

Here, we denoted this ratio as $Br$, which is (since $\Gamma(X \rightarrow e^+e^{\pm}\gamma)\ll \Gamma(X \rightarrow e^+e^{\pm})$) close to the branching ratio.

\section{Results}

Processes (1),(3),(4),(6) were simulated by making use of the CompHEP \cite{Ref42,Ref43,Ref44} and MadGraph \cite{Ref45} MC generators. Numerical results were obtained for the mass of $X$ being equal to 1000 GeV. For the presentation of the results, the relation (7) is used in differential form for photon energy spectra in both model cases ($e^+e^-$ and $e^+e^+$).
\begin{equation}
\frac{dBr_{e^+e^{\pm}\gamma}(E)}{dE}\equiv \frac{1}{\Gamma_{e^-e^{\pm}}}\frac{d\Gamma_{e^-e^{\pm}\gamma}(E)}{dE},
\end{equation}
where  $\Gamma_i$ and $Br_i$ are the widths of the respective processes and their ratio (according to (7)), and $E$ is the FSR photon energy.

\begin{figure}[H]
\begin{center}
\begin{minipage}[h]{0.496\linewidth}
\includegraphics[width=1\linewidth]{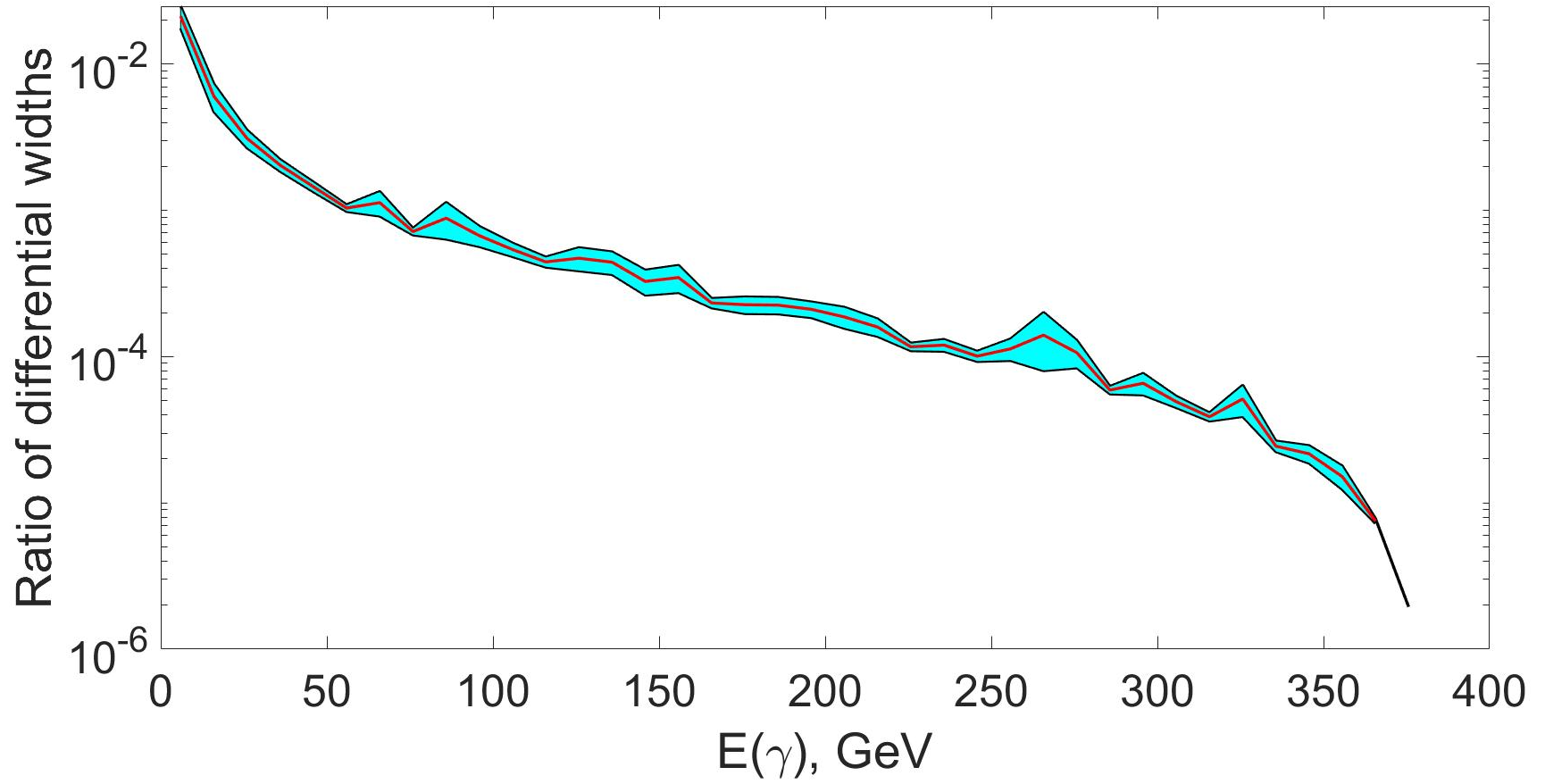}
\label{constr_10} %% метка рисунка для ссылки на него
\end{minipage}
\hfill
\begin{minipage}[h]{0.496\linewidth}
\includegraphics[width=1\linewidth]{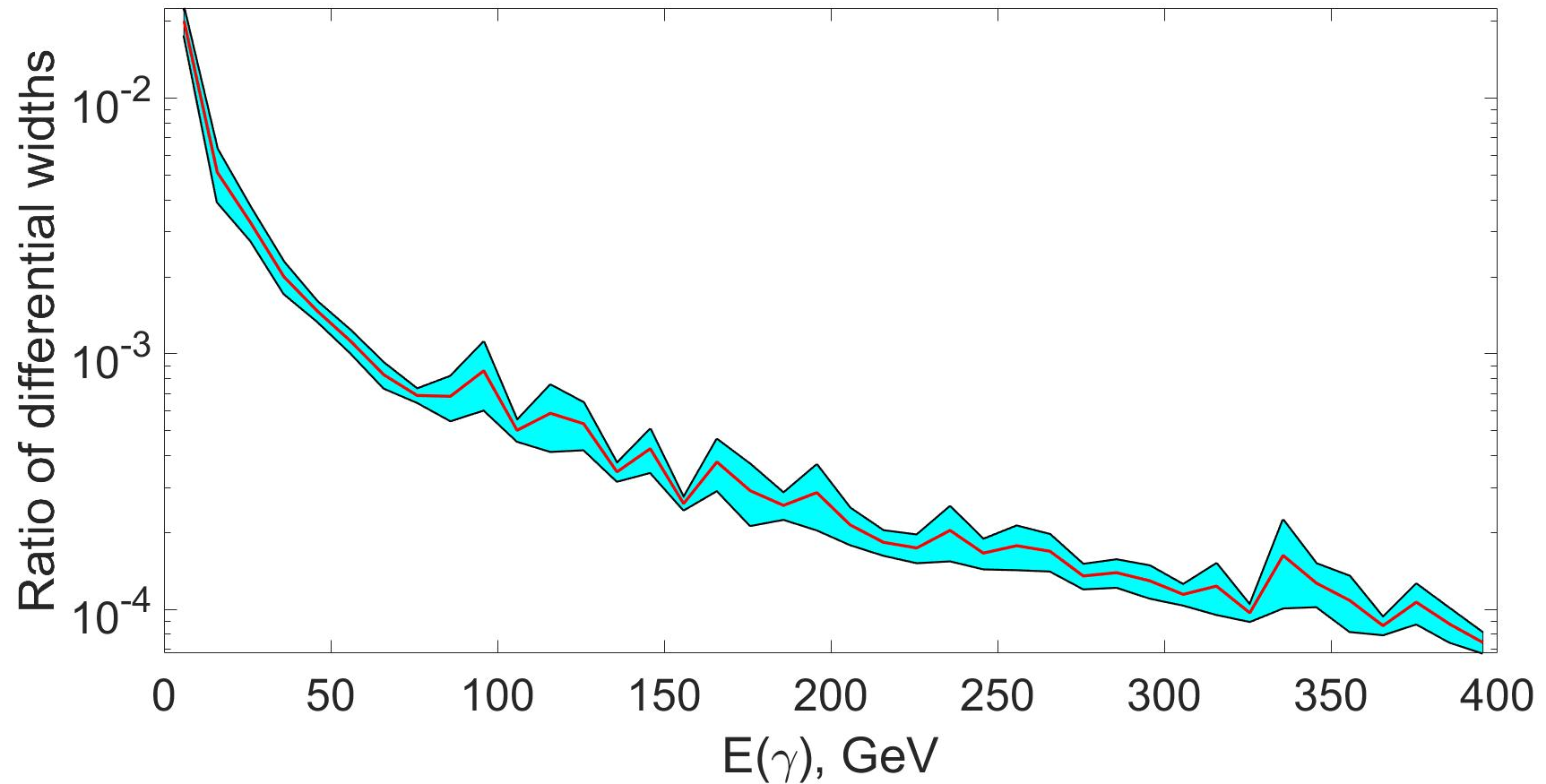}
\label{constr_01}
\end{minipage}
\end{center}
\caption{Energy distribution of photons $\frac{dBr_{e^+e^{\pm}\gamma}(E)}{dE}$ from DM particle decay through $e^+e^+$ mode (\textbf{left}) and $e^+e^-$ one (\textbf{right}). Dotted lines show errors. {Cyan color represents errors.}}
\label{fig2}
\end{figure}

\begin{figure}[H]
\begin{center}
\includegraphics[width=12.5 cm]{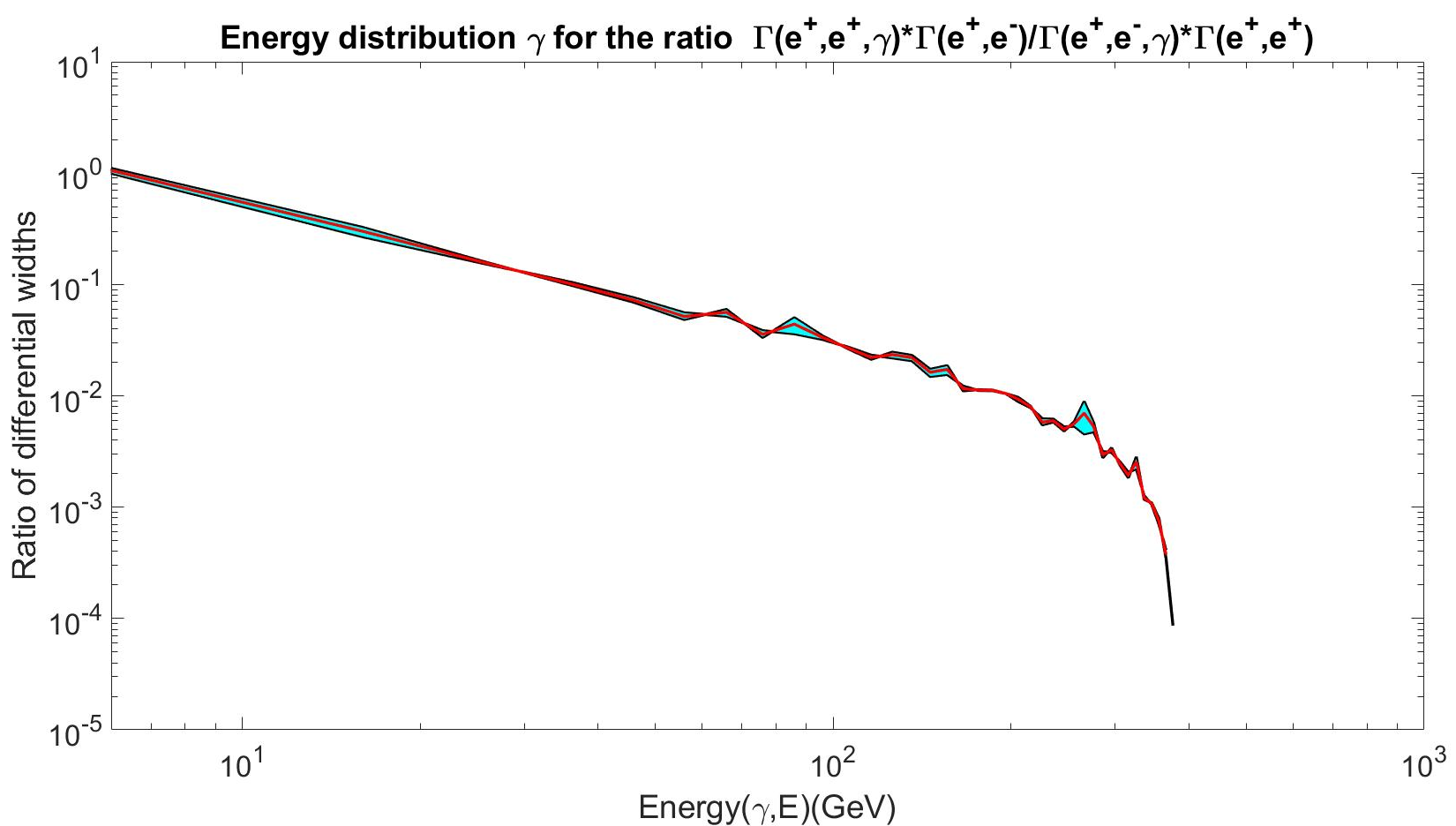}
\end{center}
\caption{The ratio $R(E)$ of photon energy spectra from the two processes $X \rightarrow e^+e^+\gamma$ and \mbox{$X \rightarrow e^+e^-\gamma$}.}
\label{fig3}
\end{figure}

The results for these two types of processes are shown in Figure \ref{fig2}. As one can note, the $X\rightarrow e^+e^-\gamma$ mode has a more smooth drop in photon energy, especially at the upper kinematic limit. This can finally be observed in Figure \ref{fig3}, where the ratio of these \mbox{two spectra }
\begin{equation}
    R(E)=\frac{dBr_{e^+e^+\gamma}(E)/dE}{dBr_{e^+e^-\gamma}(E)/dE}
\end{equation}
is shown. This is the main result, which shows the essential suppression of FSR photons in the model with the decay $X^{++}\rightarrow e^+e^+\gamma$ as compared with $X^0\rightarrow e^+e^-\gamma$ with the growth of photon energy, as was necessary for the resolution of contradiction between the DM explanation of PA and data on IGRB.

This behavior of spectra ratio has a qualitative explanation. The highest FSR photon energy corresponds to the situation when two charged leptons move with the maximum possible energy in the direction opposite to that of the photon (lepton and photon momenta are related as follows: $\vec p_{e1}=\vec p_{e2}=-\vec p_{\gamma}/2$). However, two positrons cannot be born with identical momenta because of the Pauli exclusion principle.

%\subsection{Simulation results with MadGraph \textbf{[Is this with MadGraph?]}}

\section{Conclusions}

In this work, an overview of prerequisites for solving the problem of DM explanation of positron anomaly in CR was conveyed. Such an explanation faces discrepancy with data on cosmic gamma rays. The result of this note is a suggestion of the model, which provides the suppression of FSR photons in comparison with the traditional case. The model suggested is based on a decaying doubly charged DM particle $X^{++}\rightarrow e^+e^+$. This displays the suppression of the FSR photon yield{, relatively to positrons,} for two reasons: first, we have {twice} as many positrons per photon as compared with the more conventional case $X^0\rightarrow e^+e^-$; the second, which is the main result of this note, is the effect of the suppression of FSR photons due to an identity of final charged fermions. The latter leads to an additional essential suppression of FSR. This suppression takes place in the classical case since two same charged particles do not have an electric dipole momentum and, therefore, radiation. In the quantum case, the so-called single-photon theorem tells a similar thing in an implicit way. We have shown here an effect in a specific model example that is yet to be applied to concrete astrophysical and cosmological problems. We do not show here how this suppression helps in explaining the PA problem further. This requires a separate comprehensive study. In any case, such an effect will facilitate its solution, and this is what we pay attention to.

\begin{acknowledgments} The work was performed with the financial support provided by the Russian Ministry of Science and Higher Education, project “Fundamental and applied research of cosmic rays”, No. FSWU-2023-0068. The authors would also like to thank R.Budaev, A.Kamaletdinov, A.Kirillov, M.Khlopov, M.Laletin, S.Rubin, M.Solovyov for their contribution to the work in this investigative direction or/and useful discussions.
\end{acknowledgments}

%\appendixstart
\appendix
\section[\appendixname~\thesection]{}
\label{Appendix: A}

In the minimal model of walking technicolor, there are two technical quarks, $U$ and $D$, which are being transformed by a single representation of the technicolor group $SU(2)$, and two technileptons, $\nu$' and $\zeta$. Electric charges can be chosen in the following way: +1 and 0 for $U$ and $D$ and $-$1 and $-$2 for $\nu$' and $\zeta$. 9 Goldstone bosons are produced in the model. In these models, one can implement the possibility of DM in the form of doubly charged particles. Two different cases can be considered. In the first case are an excess of $\bar{U}$$\bar{U}$  with a~charge of $-$2 and a smaller excess of $\zeta$ with a charge of +2. In this case, the main component of DM will consist of bound states of helium and $\bar{U}$$\bar{U}$: $He$$\bar{U}$$\bar{U}$. {These} are the so-called SIMPs (strongly interacting massive particles). A small component will consist in the form of bound states $\zeta$$\bar{U}$$\bar{U}$, which are the so-called WIMPs (weakly interacting massive particles). In the second case, on the other hand, an excess of $\zeta$ and a smaller excess of $UU$ are assumed. In this case, the main component of DM will consist of the states $He$$\zeta$ (SIMP) and a small component will consist of the states $UU$$\zeta$ (WIMP).

In both cases, it is assumed that $UU$ is the lightest technibaryon, and $\zeta$ is the lightest technilepton. The assumption of the smallness of the WIMP component is due to the results of underground experiments on the direct search for DM. The constraint obtained from the underground experiments requires that the relative fraction of the WIMP component has to be at the level of $\sim10^{-6}$ \cite{Ref32}. This value of the WIMP fraction and the corresponding values of initial excesses between particles and antiparticles can be obtained on the base of the mechanism of sphaleron transitions in the early universe and can be associated with an excess of baryons and leptons \cite{Ref32}. It is important to note that the $UU$ state has both charge +2 and spin 0, which is important for our final goal.

This is an example of the model, which can incorporate an unstable doubly charged DM candidate. It does not exhaust all the physical frameworks in which such candidates may be possible.
%\appendixend

% Reference section

\newpage
 \section*{References}
\bibliography{references}

%merlin.mbs apsrev4-1.bst 2010-07-25 4.21a (PWD, AO, DPC) hacked
%Control: key (0)
%Control: author (8) initials jnrlst
%Control: editor formatted (1) identically to author
%Control: production of article title (-1) disabled
%Control: page (0) single
%Control: year (1) truncated
%Control: production of eprint (0) enabled
\begin{thebibliography}{45}%
\makeatletter
\providecommand \@ifxundefined [1]{%
 \@ifx{#1\undefined}
}%
\providecommand \@ifnum [1]{%
 \ifnum #1\expandafter \@firstoftwo
 \else \expandafter \@secondoftwo
 \fi
}%
\providecommand \@ifx [1]{%
 \ifx #1\expandafter \@firstoftwo
 \else \expandafter \@secondoftwo
 \fi
}%
\providecommand \natexlab [1]{#1}%
\providecommand \enquote  [1]{``#1''}%
\providecommand \bibnamefont  [1]{#1}%
\providecommand \bibfnamefont [1]{#1}%
\providecommand \citenamefont [1]{#1}%
\providecommand \href@noop [0]{\@secondoftwo}%
\providecommand \href [0]{\begingroup \@sanitize@url \@href}%
\providecommand \@href[1]{\@@startlink{#1}\@@href}%
\providecommand \@@href[1]{\endgroup#1\@@endlink}%
\providecommand \@sanitize@url [0]{\catcode `\\12\catcode `\$12\catcode
  `\&12\catcode `\#12\catcode `\^12\catcode `\_12\catcode `\%12\relax}%
\providecommand \@@startlink[1]{}%
\providecommand \@@endlink[0]{}%
\providecommand \url  [0]{\begingroup\@sanitize@url \@url }%
\providecommand \@url [1]{\endgroup\@href {#1}{\urlprefix }}%
\providecommand \urlprefix  [0]{URL }%
\providecommand \Eprint [0]{\href }%
\providecommand \doibase [0]{http://dx.doi.org/}%
\providecommand \selectlanguage [0]{\@gobble}%
\providecommand \bibinfo  [0]{\@secondoftwo}%
\providecommand \bibfield  [0]{\@secondoftwo}%
\providecommand \translation [1]{[#1]}%
\providecommand \BibitemOpen [0]{}%
\providecommand \bibitemStop [0]{}%
\providecommand \bibitemNoStop [0]{.\EOS\space}%
\providecommand \EOS [0]{\spacefactor3000\relax}%
\providecommand \BibitemShut  [1]{\csname bibitem#1\endcsname}%
\let\auto@bib@innerbib\@empty
%</preamble>
\bibitem [{\citenamefont {Adriani}\ \emph {et~al.}(2009)\citenamefont {Adriani}
  \emph {et~al.}}]{Ref1}%
  \BibitemOpen
  \bibfield  {author} {\bibinfo {author} {\bibfnamefont {O.}~\bibnamefont
  {Adriani}} \emph {et~al.},\ }\href {\doibase 10.1038/nature07942} {\bibfield
  {journal} {\bibinfo  {journal} {Nature}\ }\textbf {\bibinfo {volume} {458}},\
  \bibinfo {pages} {607—609} (\bibinfo {year} {2009})}\BibitemShut {NoStop}%
\bibitem [{\citenamefont {{Aguilar}}\ \emph {et~al.}(2013)\citenamefont
  {{Aguilar}} \emph {et~al.}}]{Ref2}%
  \BibitemOpen
  \bibfield  {author} {\bibinfo {author} {\bibfnamefont {M.}~\bibnamefont
  {{Aguilar}}} \emph {et~al.},\ }\href {\doibase
  10.1103/PhysRevLett.110.141102} {\bibfield  {journal} {\bibinfo  {journal}
  {\prl}\ }\textbf {\bibinfo {volume} {110}},\ \bibinfo {eid} {141102}
  (\bibinfo {year} {2013})}\BibitemShut {NoStop}%
\bibitem [{\citenamefont {Ackermann}\ \emph {et~al.}(2012)\citenamefont
  {Ackermann} \emph {et~al.}}]{Ref3}%
  \BibitemOpen
  \bibfield  {author} {\bibinfo {author} {\bibfnamefont {M.}~\bibnamefont
  {Ackermann}} \emph {et~al.},\ }\href {\doibase
  10.1103/physrevlett.108.011103} {\bibfield  {journal} {\bibinfo  {journal}
  {Physical Review Letters}\ }\textbf {\bibinfo {volume} {108}} (\bibinfo
  {year} {2012}),\ 10.1103/physrevlett.108.011103}\BibitemShut {NoStop}%
\bibitem [{Ref(2017)}]{Ref4}%
  \BibitemOpen
  \href {\doibase 10.1038/nature24475} {\bibfield  {journal} {\bibinfo
  {journal} {Nature}\ }\textbf {\bibinfo {volume} {552}},\ \bibinfo {pages}
  {63} (\bibinfo {year} {2017})}\BibitemShut {NoStop}%
\bibitem [{\citenamefont {Hooper}\ \emph {et~al.}(2017)\citenamefont {Hooper},
  \citenamefont {Cholis}, \citenamefont {Linden},\ and\ \citenamefont
  {Fang}}]{Ref5}%
  \BibitemOpen
  \bibfield  {author} {\bibinfo {author} {\bibfnamefont {D.}~\bibnamefont
  {Hooper}}, \bibinfo {author} {\bibfnamefont {I.}~\bibnamefont {Cholis}},
  \bibinfo {author} {\bibfnamefont {T.}~\bibnamefont {Linden}}, \ and\ \bibinfo
  {author} {\bibfnamefont {K.}~\bibnamefont {Fang}},\ }\href {\doibase
  10.1103/physrevd.96.103013} {\bibfield  {journal} {\bibinfo  {journal}
  {Physical Review D}\ }\textbf {\bibinfo {volume} {96}} (\bibinfo {year}
  {2017}),\ 10.1103/physrevd.96.103013}\BibitemShut {NoStop}%
\bibitem [{\citenamefont {Profumo}\ \emph {et~al.}(2018)\citenamefont
  {Profumo}, \citenamefont {Reynoso-Cordova}, \citenamefont {Kaaz},\ and\
  \citenamefont {Silverman}}]{Ref6}%
  \BibitemOpen
  \bibfield  {author} {\bibinfo {author} {\bibfnamefont {S.}~\bibnamefont
  {Profumo}}, \bibinfo {author} {\bibfnamefont {J.}~\bibnamefont
  {Reynoso-Cordova}}, \bibinfo {author} {\bibfnamefont {N.}~\bibnamefont
  {Kaaz}}, \ and\ \bibinfo {author} {\bibfnamefont {M.}~\bibnamefont
  {Silverman}},\ }\href {\doibase 10.1103/physrevd.97.123008} {\bibfield
  {journal} {\bibinfo  {journal} {Physical Review D}\ }\textbf {\bibinfo
  {volume} {97}} (\bibinfo {year} {2018}),\
  10.1103/physrevd.97.123008}\BibitemShut {NoStop}%
\bibitem [{\citenamefont {Linares}\ and\ \citenamefont
  {Kachelrie{\ss}}(2021)}]{Ref7}%
  \BibitemOpen
  \bibfield  {author} {\bibinfo {author} {\bibfnamefont {M.}~\bibnamefont
  {Linares}}\ and\ \bibinfo {author} {\bibfnamefont {M.}~\bibnamefont
  {Kachelrie{\ss}}},\ }\href {\doibase 10.1088/1475-7516/2021/02/030}
  {\bibfield  {journal} {\bibinfo  {journal} {Journal of Cosmology and
  Astroparticle Physics}\ }\textbf {\bibinfo {volume} {2021}},\ \bibinfo
  {pages} {030} (\bibinfo {year} {2021})}\BibitemShut {NoStop}%
\bibitem [{\citenamefont {Orusa}\ \emph {et~al.}(2021)\citenamefont {Orusa},
  \citenamefont {Manconi}, \citenamefont {Donato},\ and\ \citenamefont
  {Mauro}}]{Ref8}%
  \BibitemOpen
  \bibfield  {author} {\bibinfo {author} {\bibfnamefont {L.}~\bibnamefont
  {Orusa}}, \bibinfo {author} {\bibfnamefont {S.}~\bibnamefont {Manconi}},
  \bibinfo {author} {\bibfnamefont {F.}~\bibnamefont {Donato}}, \ and\ \bibinfo
  {author} {\bibfnamefont {M.~D.}\ \bibnamefont {Mauro}},\ }\href {\doibase
  10.1088/1475-7516/2021/12/014} {\bibfield  {journal} {\bibinfo  {journal}
  {Journal of Cosmology and Astroparticle Physics}\ }\textbf {\bibinfo {volume}
  {2021}},\ \bibinfo {pages} {014} (\bibinfo {year} {2021})}\BibitemShut
  {NoStop}%
\bibitem [{\citenamefont {Belotsky}\ \emph {et~al.}(2020)\citenamefont
  {Belotsky}, \citenamefont {Kamaletdinov}, \citenamefont {Shlepkina},\ and\
  \citenamefont {Solovyov}}]{Ref9}%
  \BibitemOpen
  \bibfield  {author} {\bibinfo {author} {\bibfnamefont {K.~M.}\ \bibnamefont
  {Belotsky}}, \bibinfo {author} {\bibfnamefont {A.~K.}\ \bibnamefont
  {Kamaletdinov}}, \bibinfo {author} {\bibfnamefont {E.~S.}\ \bibnamefont
  {Shlepkina}}, \ and\ \bibinfo {author} {\bibfnamefont {M.~L.}\ \bibnamefont
  {Solovyov}},\ }\href {\doibase 10.3390/particles3020025} {\bibfield
  {journal} {\bibinfo  {journal} {Particles}\ }\textbf {\bibinfo {volume}
  {3}},\ \bibinfo {pages} {336} (\bibinfo {year} {2020})}\BibitemShut {NoStop}%
\bibitem [{\citenamefont {Belotsky}\ \emph
  {et~al.}(2019{\natexlab{a}})\citenamefont {Belotsky} \emph {et~al.}}]{Ref10}%
  \BibitemOpen
  \bibfield  {author} {\bibinfo {author} {\bibfnamefont {K.~M.}\ \bibnamefont
  {Belotsky}} \emph {et~al.},\ }\href {\doibase 10.1142/s0218271819410116}
  {\bibfield  {journal} {\bibinfo  {journal} {International Journal of Modern
  Physics D}\ }\textbf {\bibinfo {volume} {28}},\ \bibinfo {pages} {1941011}
  (\bibinfo {year} {2019}{\natexlab{a}})}\BibitemShut {NoStop}%
\bibitem [{\citenamefont {Belotsky}\ \emph
  {et~al.}(2017{\natexlab{a}})\citenamefont {Belotsky}, \citenamefont {Budaev},
  \citenamefont {Kirillov},\ and\ \citenamefont {Laletin}}]{Ref11}%
  \BibitemOpen
  \bibfield  {author} {\bibinfo {author} {\bibfnamefont {K.}~\bibnamefont
  {Belotsky}}, \bibinfo {author} {\bibfnamefont {R.}~\bibnamefont {Budaev}},
  \bibinfo {author} {\bibfnamefont {A.}~\bibnamefont {Kirillov}}, \ and\
  \bibinfo {author} {\bibfnamefont {M.}~\bibnamefont {Laletin}},\ }\href
  {\doibase 10.1088/1475-7516/2017/01/021} {\bibfield  {journal} {\bibinfo
  {journal} {Journal of Cosmology and Astroparticle Physics}\ }\textbf
  {\bibinfo {volume} {2017}},\ \bibinfo {pages} {021} (\bibinfo {year}
  {2017}{\natexlab{a}})}\BibitemShut {NoStop}%
\bibitem [{\citenamefont {Belotsky}\ \emph
  {et~al.}(2019{\natexlab{b}})\citenamefont {Belotsky}, \citenamefont
  {Kamaletdinov}, \citenamefont {Laletin},\ and\ \citenamefont
  {Solovyov}}]{Ref12}%
  \BibitemOpen
  \bibfield  {author} {\bibinfo {author} {\bibfnamefont {K.}~\bibnamefont
  {Belotsky}}, \bibinfo {author} {\bibfnamefont {A.}~\bibnamefont
  {Kamaletdinov}}, \bibinfo {author} {\bibfnamefont {M.}~\bibnamefont
  {Laletin}}, \ and\ \bibinfo {author} {\bibfnamefont {M.}~\bibnamefont
  {Solovyov}},\ }\href {\doibase 10.1016/j.dark.2019.100333} {\bibfield
  {journal} {\bibinfo  {journal} {Physics of the Dark Universe}\ }\textbf
  {\bibinfo {volume} {26}},\ \bibinfo {pages} {100333} (\bibinfo {year}
  {2019}{\natexlab{b}})}\BibitemShut {NoStop}%
\bibitem [{\citenamefont {Huang}\ \emph {et~al.}(2020)\citenamefont {Huang},
  \citenamefont {Liu}, \citenamefont {Joshi},\ and\ \citenamefont
  {Wang}}]{Ref13}%
  \BibitemOpen
  \bibfield  {author} {\bibinfo {author} {\bibfnamefont {Z.-Q.}\ \bibnamefont
  {Huang}}, \bibinfo {author} {\bibfnamefont {R.-Y.}\ \bibnamefont {Liu}},
  \bibinfo {author} {\bibfnamefont {J.~C.}\ \bibnamefont {Joshi}}, \ and\
  \bibinfo {author} {\bibfnamefont {X.-Y.}\ \bibnamefont {Wang}},\ }\href
  {\doibase 10.3847/1538-4357/ab88cb} {\bibfield  {journal} {\bibinfo
  {journal} {The Astrophysical Journal}\ }\textbf {\bibinfo {volume} {895}},\
  \bibinfo {pages} {53} (\bibinfo {year} {2020})}\BibitemShut {NoStop}%
\bibitem [{\citenamefont {Yang}\ \emph {et~al.}(2017)\citenamefont {Yang},
  \citenamefont {Su},\ and\ \citenamefont {Zhao}}]{Ref14}%
  \BibitemOpen
  \bibfield  {author} {\bibinfo {author} {\bibfnamefont {F.}~\bibnamefont
  {Yang}}, \bibinfo {author} {\bibfnamefont {M.}~\bibnamefont {Su}}, \ and\
  \bibinfo {author} {\bibfnamefont {Y.}~\bibnamefont {Zhao}},\ }\href@noop {}
  {\enquote {\bibinfo {title} {Dark matter annihilation from nearby
  ultra-compact micro halos to explain the tentative excess at ~1.4 tev in
  dampe data},}\ } (\bibinfo {year} {2017}),\ \Eprint
  {http://arxiv.org/abs/1712.01724} {arXiv:1712.01724 [astro-ph.HE]}
  \BibitemShut {NoStop}%
\bibitem [{\citenamefont {Cheng}\ \emph {et~al.}(2020)\citenamefont {Cheng}
  \emph {et~al.}}]{Ref15}%
  \BibitemOpen
  \bibfield  {author} {\bibinfo {author} {\bibfnamefont {J.-G.}\ \bibnamefont
  {Cheng}} \emph {et~al.},\ }\href {\doibase 10.1093/mnras/staa2092} {\bibfield
   {journal} {\bibinfo  {journal} {Monthly Notices of the Royal Astronomical
  Society}\ }\textbf {\bibinfo {volume} {497}},\ \bibinfo {pages} {2486}
  (\bibinfo {year} {2020})}\BibitemShut {NoStop}%
\bibitem [{\citenamefont {Ge}\ \emph {et~al.}(2020)\citenamefont {Ge},
  \citenamefont {He}, \citenamefont {Wang},\ and\ \citenamefont
  {Yuan}}]{Ref16}%
  \BibitemOpen
  \bibfield  {author} {\bibinfo {author} {\bibfnamefont {S.-F.}\ \bibnamefont
  {Ge}}, \bibinfo {author} {\bibfnamefont {H.-J.}\ \bibnamefont {He}}, \bibinfo
  {author} {\bibfnamefont {Y.-C.}\ \bibnamefont {Wang}}, \ and\ \bibinfo
  {author} {\bibfnamefont {Q.}~\bibnamefont {Yuan}},\ }\href {\doibase
  10.1016/j.nuclphysb.2020.115140} {\bibfield  {journal} {\bibinfo  {journal}
  {Nuclear Physics B}\ }\textbf {\bibinfo {volume} {959}},\ \bibinfo {pages}
  {115140} (\bibinfo {year} {2020})}\BibitemShut {NoStop}%
\bibitem [{\citenamefont {Chen}\ \emph {et~al.}(2015)\citenamefont {Chen},
  \citenamefont {Chiang},\ and\ \citenamefont {Nomura}}]{Ref17}%
  \BibitemOpen
  \bibfield  {author} {\bibinfo {author} {\bibfnamefont {C.-H.}\ \bibnamefont
  {Chen}}, \bibinfo {author} {\bibfnamefont {C.-W.}\ \bibnamefont {Chiang}}, \
  and\ \bibinfo {author} {\bibfnamefont {T.}~\bibnamefont {Nomura}},\ }\href
  {\doibase 10.1016/j.physletb.2015.06.035} {\bibfield  {journal} {\bibinfo
  {journal} {Physics Letters B}\ }\textbf {\bibinfo {volume} {747}},\ \bibinfo
  {pages} {495} (\bibinfo {year} {2015})}\BibitemShut {NoStop}%
\bibitem [{\citenamefont {Diamanti}\ \emph {et~al.}(2014)\citenamefont
  {Diamanti} \emph {et~al.}}]{Ref18}%
  \BibitemOpen
  \bibfield  {author} {\bibinfo {author} {\bibfnamefont {R.}~\bibnamefont
  {Diamanti}} \emph {et~al.},\ }\href {\doibase 10.1088/1475-7516/2014/02/017}
  {\bibfield  {journal} {\bibinfo  {journal} {Journal of Cosmology and
  Astroparticle Physics}\ }\textbf {\bibinfo {volume} {2014}},\ \bibinfo
  {pages} {017} (\bibinfo {year} {2014})}\BibitemShut {NoStop}%
\bibitem [{\citenamefont {Xiang}\ \emph {et~al.}(2017)\citenamefont {Xiang},
  \citenamefont {Bi}, \citenamefont {Lin},\ and\ \citenamefont {Yin}}]{Ref19}%
  \BibitemOpen
  \bibfield  {author} {\bibinfo {author} {\bibfnamefont {Q.-F.}\ \bibnamefont
  {Xiang}}, \bibinfo {author} {\bibfnamefont {X.-J.}\ \bibnamefont {Bi}},
  \bibinfo {author} {\bibfnamefont {S.-J.}\ \bibnamefont {Lin}}, \ and\
  \bibinfo {author} {\bibfnamefont {P.-F.}\ \bibnamefont {Yin}},\ }\href
  {\doibase 10.1016/j.physletb.2017.09.003} {\bibfield  {journal} {\bibinfo
  {journal} {Physics Letters B}\ }\textbf {\bibinfo {volume} {773}},\ \bibinfo
  {pages} {448} (\bibinfo {year} {2017})}\BibitemShut {NoStop}%
\bibitem [{\citenamefont {Fang}\ \emph {et~al.}(2018)\citenamefont {Fang},
  \citenamefont {Bi},\ and\ \citenamefont {Yin}}]{Ref20}%
  \BibitemOpen
  \bibfield  {author} {\bibinfo {author} {\bibfnamefont {K.}~\bibnamefont
  {Fang}}, \bibinfo {author} {\bibfnamefont {X.-J.}\ \bibnamefont {Bi}}, \ and\
  \bibinfo {author} {\bibfnamefont {P.-F.}\ \bibnamefont {Yin}},\ }\href
  {\doibase 10.3847/1538-4357/aaa710} {\bibfield  {journal} {\bibinfo
  {journal} {The Astrophysical Journal}\ }\textbf {\bibinfo {volume} {854}},\
  \bibinfo {pages} {57} (\bibinfo {year} {2018})}\BibitemShut {NoStop}%
\bibitem [{\citenamefont {Kachelrie{\ss}}\ \emph {et~al.}(2018)\citenamefont
  {Kachelrie{\ss}}, \citenamefont {Neronov},\ and\ \citenamefont
  {Semikoz}}]{Ref21}%
  \BibitemOpen
  \bibfield  {author} {\bibinfo {author} {\bibfnamefont {M.}~\bibnamefont
  {Kachelrie{\ss}}}, \bibinfo {author} {\bibfnamefont {A.}~\bibnamefont
  {Neronov}}, \ and\ \bibinfo {author} {\bibfnamefont {D.}~\bibnamefont
  {Semikoz}},\ }\href {\doibase 10.1103/physrevd.97.063011} {\bibfield
  {journal} {\bibinfo  {journal} {Physical Review D}\ }\textbf {\bibinfo
  {volume} {97}} (\bibinfo {year} {2018}),\
  10.1103/physrevd.97.063011}\BibitemShut {NoStop}%
\bibitem [{\citenamefont {Blum}\ \emph {et~al.}(2013)\citenamefont {Blum},
  \citenamefont {Katz},\ and\ \citenamefont {Waxman}}]{Ref22}%
  \BibitemOpen
  \bibfield  {author} {\bibinfo {author} {\bibfnamefont {K.}~\bibnamefont
  {Blum}}, \bibinfo {author} {\bibfnamefont {B.}~\bibnamefont {Katz}}, \ and\
  \bibinfo {author} {\bibfnamefont {E.}~\bibnamefont {Waxman}},\ }\href
  {\doibase 10.1103/PhysRevLett.111.211101} {\bibfield  {journal} {\bibinfo
  {journal} {Physical review letters}\ }\textbf {\bibinfo {volume} {111}},\
  \bibinfo {pages} {211101} (\bibinfo {year} {2013})}\BibitemShut {NoStop}%
\bibitem [{\citenamefont {Tomassetti}(2015)}]{Ref23}%
  \BibitemOpen
  \bibfield  {author} {\bibinfo {author} {\bibfnamefont {N.}~\bibnamefont
  {Tomassetti}},\ }\href {\doibase 10.1103/PhysRevD.92.081301} {\bibfield
  {journal} {\bibinfo  {journal} {Physical Review D}\ }\textbf {\bibinfo
  {volume} {92}} (\bibinfo {year} {2015}),\
  10.1103/PhysRevD.92.081301}\BibitemShut {NoStop}%
\bibitem [{\citenamefont {Kappl}\ and\ \citenamefont {Reinert}(2017)}]{Ref24}%
  \BibitemOpen
  \bibfield  {author} {\bibinfo {author} {\bibfnamefont {R.}~\bibnamefont
  {Kappl}}\ and\ \bibinfo {author} {\bibfnamefont {A.}~\bibnamefont
  {Reinert}},\ }\href@noop {} {\enquote {\bibinfo {title} {Secondary cosmic
  positrons in an inhomogeneous diffusion model},}\ } (\bibinfo {year}
  {2017}),\ \Eprint {http://arxiv.org/abs/1609.01300} {arXiv:1609.01300
  [astro-ph.HE]} \BibitemShut {NoStop}%
\bibitem [{\citenamefont {Aguilar}\ \emph {et~al.}(2021)\citenamefont {Aguilar}
  \emph {et~al.}}]{Ref25}%
  \BibitemOpen
  \bibfield  {author} {\bibinfo {author} {\bibfnamefont {M.}~\bibnamefont
  {Aguilar}} \emph {et~al.} (\bibinfo {collaboration} {AMS}),\ }\href {\doibase
  10.1016/j.physrep.2020.09.003} {\bibfield  {journal} {\bibinfo  {journal}
  {Phys. Rept.}\ }\textbf {\bibinfo {volume} {894}},\ \bibinfo {pages} {1}
  (\bibinfo {year} {2021})}\BibitemShut {NoStop}%
\bibitem [{\citenamefont {Belotsky}\ \emph
  {et~al.}(2017{\natexlab{b}})\citenamefont {Belotsky}, \citenamefont {Budaev},
  \citenamefont {Kirillov},\ and\ \citenamefont {Solovyov}}]{Ref26}%
  \BibitemOpen
  \bibfield  {author} {\bibinfo {author} {\bibfnamefont {K.~M.}\ \bibnamefont
  {Belotsky}}, \bibinfo {author} {\bibfnamefont {R.~I.}\ \bibnamefont
  {Budaev}}, \bibinfo {author} {\bibfnamefont {A.~A.}\ \bibnamefont
  {Kirillov}}, \ and\ \bibinfo {author} {\bibfnamefont {M.~L.}\ \bibnamefont
  {Solovyov}},\ }\href {\doibase 10.1088/1742-6596/798/1/012084} {\bibfield
  {journal} {\bibinfo  {journal} {J. Phys. Conf. Ser.}\ }\textbf {\bibinfo
  {volume} {798}},\ \bibinfo {pages} {012084} (\bibinfo {year}
  {2017}{\natexlab{b}})}\BibitemShut {NoStop}%
\bibitem [{\citenamefont {Alekseev}\ \emph
  {et~al.}(2016{\natexlab{a}})\citenamefont {Alekseev} \emph {et~al.}}]{Ref27}%
  \BibitemOpen
  \bibfield  {author} {\bibinfo {author} {\bibfnamefont {V.}~\bibnamefont
  {Alekseev}} \emph {et~al.},\ }\href {\doibase 10.1088/1742-6596/675/1/012023}
  {\bibfield  {journal} {\bibinfo  {journal} {Journal of Physics: Conference
  Series}\ }\textbf {\bibinfo {volume} {675}},\ \bibinfo {pages} {012023}
  (\bibinfo {year} {2016}{\natexlab{a}})}\BibitemShut {NoStop}%
\bibitem [{\citenamefont {Alekseev}\ \emph
  {et~al.}(2016{\natexlab{b}})\citenamefont {Alekseev} \emph {et~al.}}]{Ref28}%
  \BibitemOpen
  \bibfield  {author} {\bibinfo {author} {\bibfnamefont {V.}~\bibnamefont
  {Alekseev}} \emph {et~al.},\ }\href {\doibase 10.1088/1742-6596/675/1/012026}
  {\bibfield  {journal} {\bibinfo  {journal} {Journal of Physics: Conference
  Series}\ }\textbf {\bibinfo {volume} {675}},\ \bibinfo {pages} {012026}
  (\bibinfo {year} {2016}{\natexlab{b}})}\BibitemShut {NoStop}%
\bibitem [{\citenamefont {{Alekseev}}\ \emph {et~al.}(2017)\citenamefont
  {{Alekseev}} \emph {et~al.}}]{Ref29}%
  \BibitemOpen
  \bibfield  {author} {\bibinfo {author} {\bibfnamefont {V.~V.}\ \bibnamefont
  {{Alekseev}}} \emph {et~al.},\ }\href {\doibase 10.1134/S1063778817040020}
  {\bibfield  {journal} {\bibinfo  {journal} {Physics of Atomic Nuclei}\
  }\textbf {\bibinfo {volume} {80}},\ \bibinfo {pages} {713} (\bibinfo {year}
  {2017})}\BibitemShut {NoStop}%
\bibitem [{\citenamefont {Belotsky}\ \emph {et~al.}(2018)\citenamefont
  {Belotsky}, \citenamefont {Kirillov},\ and\ \citenamefont
  {Solovyov}}]{Ref30}%
  \BibitemOpen
  \bibfield  {author} {\bibinfo {author} {\bibfnamefont {K.~M.}\ \bibnamefont
  {Belotsky}}, \bibinfo {author} {\bibfnamefont {A.~A.}\ \bibnamefont
  {Kirillov}}, \ and\ \bibinfo {author} {\bibfnamefont {M.~L.}\ \bibnamefont
  {Solovyov}},\ }\href {\doibase 10.1142/s0218271818410109} {\bibfield
  {journal} {\bibinfo  {journal} {International Journal of Modern Physics D}\
  }\textbf {\bibinfo {volume} {27}},\ \bibinfo {pages} {1841010} (\bibinfo
  {year} {2018})}\BibitemShut {NoStop}%
\bibitem [{\citenamefont {Solovyov}\ \emph {et~al.}(2020)\citenamefont
  {Solovyov}, \citenamefont {Rakhimova},\ and\ \citenamefont
  {Belotsky}}]{Ref31}%
  \BibitemOpen
  \bibfield  {author} {\bibinfo {author} {\bibfnamefont {M.}~\bibnamefont
  {Solovyov}}, \bibinfo {author} {\bibfnamefont {M.}~\bibnamefont {Rakhimova}},
  \ and\ \bibinfo {author} {\bibfnamefont {K.}~\bibnamefont {Belotsky}},\
  }\href@noop {} {\bibfield  {journal} {\bibinfo  {journal} {Proceedings of the
  23rd Bled Workshop “What Comes Beyond Standard Models?”, Bled, Slovenia}\
  }\textbf {\bibinfo {volume} {21}},\ \bibinfo {pages} {156} (\bibinfo {year}
  {2020})}\BibitemShut {NoStop}%
\bibitem [{\citenamefont {Belotsky}\ \emph {et~al.}(2014)\citenamefont
  {Belotsky}, \citenamefont {Khlopov}, \citenamefont {Kouvaris},\ and\
  \citenamefont {Laletin}}]{Ref32}%
  \BibitemOpen
  \bibfield  {author} {\bibinfo {author} {\bibfnamefont {K.}~\bibnamefont
  {Belotsky}}, \bibinfo {author} {\bibfnamefont {M.}~\bibnamefont {Khlopov}},
  \bibinfo {author} {\bibfnamefont {C.}~\bibnamefont {Kouvaris}}, \ and\
  \bibinfo {author} {\bibfnamefont {M.}~\bibnamefont {Laletin}},\ }\href
  {\doibase 10.1155/2014/214258} {\bibfield  {journal} {\bibinfo  {journal}
  {Advances in High Energy Physics}\ }\textbf {\bibinfo {volume} {2014}},\
  \bibinfo {pages} {1} (\bibinfo {year} {2014})}\BibitemShut {NoStop}%
\bibitem [{\citenamefont {Belotsky}\ \emph {et~al.}(2015)\citenamefont
  {Belotsky}, \citenamefont {Khlopov}, \citenamefont {Kouvaris},\ and\
  \citenamefont {Laletin}}]{Ref33}%
  \BibitemOpen
  \bibfield  {author} {\bibinfo {author} {\bibfnamefont {K.}~\bibnamefont
  {Belotsky}}, \bibinfo {author} {\bibfnamefont {M.}~\bibnamefont {Khlopov}},
  \bibinfo {author} {\bibfnamefont {C.}~\bibnamefont {Kouvaris}}, \ and\
  \bibinfo {author} {\bibfnamefont {M.}~\bibnamefont {Laletin}},\ }\href
  {\doibase 10.1142/s0218271815450042} {\bibfield  {journal} {\bibinfo
  {journal} {International Journal of Modern Physics D}\ }\textbf {\bibinfo
  {volume} {24}},\ \bibinfo {pages} {1545004} (\bibinfo {year}
  {2015})}\BibitemShut {NoStop}%
\bibitem [{\citenamefont {Khlopov}(2011)}]{Ref34}%
  \BibitemOpen
  \bibfield  {author} {\bibinfo {author} {\bibfnamefont {M.}~\bibnamefont
  {Khlopov}},\ }\href {\doibase 10.1142/S0217732311037194} {\bibfield
  {journal} {\bibinfo  {journal} {Modern Physics Letters A}\ }\textbf {\bibinfo
  {volume} {26}} (\bibinfo {year} {2011}),\
  10.1142/S0217732311037194}\BibitemShut {NoStop}%
\bibitem [{\citenamefont {Khlopov}(2005)}]{Ref35}%
  \BibitemOpen
  \bibfield  {author} {\bibinfo {author} {\bibfnamefont {M.~Y.}\ \bibnamefont
  {Khlopov}},\ }\href@noop {} {\enquote {\bibinfo {title} {Composite dark
  matter from 4th generation},}\ } (\bibinfo {year} {2005}),\ \Eprint
  {http://arxiv.org/abs/astro-ph/0511796} {arXiv:astro-ph/0511796 [astro-ph]}
  \BibitemShut {NoStop}%
\bibitem [{\citenamefont {Khlopov}\ and\ \citenamefont
  {Kouvaris}(2008)}]{Ref36}%
  \BibitemOpen
  \bibfield  {author} {\bibinfo {author} {\bibfnamefont {M.~Y.}\ \bibnamefont
  {Khlopov}}\ and\ \bibinfo {author} {\bibfnamefont {C.}~\bibnamefont
  {Kouvaris}},\ }\href {\doibase 10.1103/physrevd.78.065040} {\bibfield
  {journal} {\bibinfo  {journal} {Physical Review D}\ }\textbf {\bibinfo
  {volume} {78}} (\bibinfo {year} {2008}),\
  10.1103/physrevd.78.065040}\BibitemShut {NoStop}%
\bibitem [{\citenamefont {Khlopov}\ \emph {et~al.}(2006)\citenamefont
  {Khlopov}, \citenamefont {Stephan},\ and\ \citenamefont {Fargion}}]{Ref37}%
  \BibitemOpen
  \bibfield  {author} {\bibinfo {author} {\bibfnamefont {M.~Y.}\ \bibnamefont
  {Khlopov}}, \bibinfo {author} {\bibfnamefont {C.~A.}\ \bibnamefont
  {Stephan}}, \ and\ \bibinfo {author} {\bibfnamefont {D.}~\bibnamefont
  {Fargion}},\ }\href {\doibase 10.1088/0264-9381/23/24/008} {\bibfield
  {journal} {\bibinfo  {journal} {Classical and Quantum Gravity}\ }\textbf
  {\bibinfo {volume} {23}},\ \bibinfo {pages} {7305} (\bibinfo {year}
  {2006})}\BibitemShut {NoStop}%
\bibitem [{\citenamefont {Beylin}\ \emph {et~al.}(2019)\citenamefont {Beylin},
  \citenamefont {Khlopov}, \citenamefont {Kuksa},\ and\ \citenamefont
  {Volchanskiy}}]{Ref38}%
  \BibitemOpen
  \bibfield  {author} {\bibinfo {author} {\bibfnamefont {V.}~\bibnamefont
  {Beylin}}, \bibinfo {author} {\bibfnamefont {M.}~\bibnamefont {Khlopov}},
  \bibinfo {author} {\bibfnamefont {V.}~\bibnamefont {Kuksa}}, \ and\ \bibinfo
  {author} {\bibfnamefont {N.}~\bibnamefont {Volchanskiy}},\ }\href {\doibase
  10.3390/sym11040587} {\bibfield  {journal} {\bibinfo  {journal} {Symmetry}\
  }\textbf {\bibinfo {volume} {11}},\ \bibinfo {pages} {587} (\bibinfo {year}
  {2019})}\BibitemShut {NoStop}%
\bibitem [{\citenamefont {Li}\ \emph {et~al.}(2023)\citenamefont {Li},
  \citenamefont {Liu},\ and\ \citenamefont {Xue}}]{Ref39}%
  \BibitemOpen
  \bibfield  {author} {\bibinfo {author} {\bibfnamefont {Y.}~\bibnamefont
  {Li}}, \bibinfo {author} {\bibfnamefont {Z.}~\bibnamefont {Liu}}, \ and\
  \bibinfo {author} {\bibfnamefont {Y.}~\bibnamefont {Xue}},\ }\href {\doibase
  10.1088/1475-7516/2023/05/060} {\bibfield  {journal} {\bibinfo  {journal}
  {Journal of Cosmology and Astroparticle Physics}\ }\textbf {\bibinfo {volume}
  {2023}},\ \bibinfo {pages} {060} (\bibinfo {year} {2023})}\BibitemShut
  {NoStop}%
\bibitem [{\citenamefont {Fedderke}\ \emph {et~al.}(2020)\citenamefont
  {Fedderke}, \citenamefont {Graham},\ and\ \citenamefont {Rajendran}}]{Ref40}%
  \BibitemOpen
  \bibfield  {author} {\bibinfo {author} {\bibfnamefont {M.}~\bibnamefont
  {Fedderke}}, \bibinfo {author} {\bibfnamefont {P.}~\bibnamefont {Graham}}, \
  and\ \bibinfo {author} {\bibfnamefont {S.}~\bibnamefont {Rajendran}},\ }\href
  {\doibase 10.1103/PhysRevD.101.115021} {\bibfield  {journal} {\bibinfo
  {journal} {Physical Review D}\ }\textbf {\bibinfo {volume} {101}} (\bibinfo
  {year} {2020}),\ 10.1103/PhysRevD.101.115021}\BibitemShut {NoStop}%
\bibitem [{\citenamefont {Brown}(1995)}]{Ref41}%
  \BibitemOpen
  \bibfield  {author} {\bibinfo {author} {\bibfnamefont {R.~W.}\ \bibnamefont
  {Brown}},\ }\href {\doibase 10.1063/1.49310} {\bibfield  {journal} {\bibinfo
  {journal} {AIP Conf. Proc.}\ }\textbf {\bibinfo {volume} {350}},\ \bibinfo
  {pages} {261} (\bibinfo {year} {1995})},\ \Eprint
  {http://arxiv.org/abs/hep-th/9506018} {arXiv:hep-th/9506018} \BibitemShut
  {NoStop}%
\bibitem [{\citenamefont {Boos}\ \emph {et~al.}(2004)\citenamefont {Boos} \emph
  {et~al.}}]{Ref42}%
  \BibitemOpen
  \bibfield  {author} {\bibinfo {author} {\bibfnamefont {E.}~\bibnamefont
  {Boos}} \emph {et~al.},\ }\href {\doibase 10.1016/j.nima.2004.07.096}
  {\bibfield  {journal} {\bibinfo  {journal} {Nuclear Instruments and Methods
  in Physics Research Section A: Accelerators, Spectrometers, Detectors and
  Associated Equipment}\ }\textbf {\bibinfo {volume} {534}},\ \bibinfo {pages}
  {250} (\bibinfo {year} {2004})}\BibitemShut {NoStop}%
\bibitem [{\citenamefont {Pukhov}\ \emph {et~al.}(2000)\citenamefont {Pukhov}
  \emph {et~al.}}]{Ref43}%
  \BibitemOpen
  \bibfield  {author} {\bibinfo {author} {\bibfnamefont {A.}~\bibnamefont
  {Pukhov}} \emph {et~al.},\ }\href@noop {} {\enquote {\bibinfo {title}
  {Comphep - a package for evaluation of feynman diagrams and integration over
  multi-particle phase space. user's manual for version 33},}\ } (\bibinfo
  {year} {2000}),\ \Eprint {http://arxiv.org/abs/hep-ph/9908288}
  {arXiv:hep-ph/9908288 [hep-ph]} \BibitemShut {NoStop}%
\bibitem [{Ref()}]{Ref44}%
  \BibitemOpen
  \href@noop {} {\enquote {\bibinfo {title} {Comphep website},}\ }\bibinfo
  {howpublished} {\url{http://comphep.sinp.msu.ru}},\ \bibinfo {note}
  {accessed: 2021-09-05}\BibitemShut {NoStop}%
\bibitem [{\citenamefont {Alwall}\ \emph {et~al.}(2014)\citenamefont {Alwall}
  \emph {et~al.}}]{Ref45}%
  \BibitemOpen
  \bibfield  {author} {\bibinfo {author} {\bibfnamefont {J.}~\bibnamefont
  {Alwall}} \emph {et~al.},\ }\href {\doibase 10.1007/jhep07(2014)079}
  {\bibfield  {journal} {\bibinfo  {journal} {Journal of High Energy Physics}\
  }\textbf {\bibinfo {volume} {2014}} (\bibinfo {year} {2014}),\
  10.1007/jhep07(2014)079}\BibitemShut {NoStop}%
\end{thebibliography}%

\end{document}